\documentclass[12pt,leqno,a4paper]{article}
\usepackage{amsmath,amsthm,enumerate}
\usepackage[latin1]{inputenc}
\usepackage[T1]{fontenc}
\usepackage[english]{babel}


\newcommand*\proofnamestyle{\itshape}

\DeclareMathOperator{\tr}{Tr}

\begin{document}

    \title{The Wigner-Yanase entropy\\
     is not subadditive}
      \author{Frank Hansen}
      \date{September 7, 2006\\
      {\tiny Revised November 20, 2006}}

      \maketitle

      \begin{abstract}
      Wigner and Yanase introduced in 1963 the Wigner-Yanase entropy defined as minus the skew information of
      a state with respect to a conserved observable. They proved that the Wigner-Yanase entropy is a concave
      function in the state and conjectured that it is subadditive with respect to the aggregation of possibly interacting subsystems.
      While this turned out to be true for the quantum-mechanical entropy, we negate the conjecture for the
      Wigner-Yanase entropy by providing a counter example.
      \end{abstract}

    \section{Introduction}

    The Wigner-Yanase-Dyson entropy $ S_p(\rho, k) $ is defined \cite{kn:wigner:1963} by setting
    \[
    S_p(\rho, k)=\frac{1}{2} \tr[\rho^p, k][\rho^{1-p}, k]\qquad 0<p<1,
    \]
    where $ \rho $ is a state (or density matrix) and $ k $ is a conserved (self-adjoint) observable. 
    Since $ S_p(\rho, k) $ is homogeneous in $ \rho $
    we may regard it as a function defined on the set of all positive definite operators on a Hilbert space $ H. $
    If the dimension of $ H $ is infinite, we usually impose the condition that $ k $ is a trace class operator.
    Note also that $ S_p(\rho, k) $ is minus the Wigner-Yanase-Dyson
    skew information in the state $ \rho $ with respect to the (conserved) observable $ k. $

    Wigner and Yanase proved concavity in the state $ \rho $ of the Wigner-Yanase entropy
    \[
    S(\rho, k)=\frac{1}{2} \tr[\rho^{1/2}, k]^2
    \]
    and conjectured that also the  Wigner-Yanase-Dyson entropy is concave in $ \rho $ for $ 0<p<1. $ The conjecture was
    eventually proved by Lieb \cite{kn:lieb:1973:1}.
    The result is in line with the well known concavity of the quantum-mechanical entropy
    \[
    S(\rho)=-\tr\rho\log\rho.
    \]
    Wigner and Yanase  also conjectured \cite{kn:wigner:1963} that the Wigner-Yanase entropy is subadditive in the
    following sense. Let $ H_{12}=H_1\otimes H_2 $ be a tensor product of two Hilbert spaces $ H_1 $ and $ H_2, $
    and let $ \rho_{12} $
    be a positive definite operator on $ H_{12}. $ In physical applications one also
    requires that $ \tr\rho_{12}=1. $ The partial trace $ \rho_1=\tr_1\rho_{12} $ is the operator on $ H_1 $
    defined by setting
    \[
    (\xi\mid\rho_1\eta)=\sum_{i\in I} (\xi\otimes e_i\mid\rho_{12} (\eta\otimes e_i))\qquad\xi,\eta\in H_1
    \]
    where $ (e_i)_{i\in I} $ is any orthonormal basis in $ H_2. $ We similarly define the partial trace
    $ \rho_2=\tr_2\rho_{12} $ on $ H_2. $ Let $ k_1 $ (respectively $ k_2) $ be a self-adjoint operator on
    $ H_1 $ (respectively $ H_2) $ and define
    \[
    k_{12}=k_1\otimes 1_2 + 1_1\otimes k_2.
    \]
    Subadditivity of the Wigner-Yanase-Dyson entropy is the condition
    \begin{equation}\label{subadditivity of the WY-entropy}
    S_p(\rho_{12}, k_{12})\le S_p(\rho_1, k_1)+S_p(\rho_2, k_2).
    \end{equation}
    Wigner and Yanase proved subadditivity (\ref{subadditivity of the WY-entropy})
    if $ \rho_{12}=\rho_1\otimes\rho_2 $ is a simple tensor of density matrices
    $ \rho_1 $ and $ \rho_2 $ (with equality),
    or if $ \rho_{12} $ is a pure state \cite{kn:wigner:1963}. Lieb also noted \cite{kn:lieb:1973:1} that
    (\ref{subadditivity of the WY-entropy}) holds if $ k_1=0 $ or $ k_2=0. $
    
    The notion of subadditivity for the classical entropy with respect to the aggregation
    of not necessarily isolated systems goes back to Gibbs \cite{kn:gibbs} and were later used by
    Kolmogorov and Sinai. A historical account may
    be found in Werhl \cite{kn:wehrl:1978}, cf. also Ruelle \cite[Proposition 7.2.6]{kn:ruelle:1969}.
    It was therefore natural to expect the same properties of the
    quantum-mechanical entropy. Subadditivity of the quantum-mechanical entropy was proved by
    Landford and Robinson \cite{kn:lanford:1968} who cited earlier partial results by Delbrück and Molière,
    and R. Jost, cf. also Araki and Lieb \cite{kn:araki:1970}, 
    while strong subadditivity were conjectured by Landford and Robinson \cite{kn:lanford:1968}
    and proved by Lieb and Ruskai \cite{kn:lieb:1973:3}, cf. also \cite{kn:nielsen:2005}.

    The notion of strong subadditivity for the Wigner-Yanase-Dyson entropy is defined in the following way.
    Let $ H_{123}=H_1\otimes H_2\otimes H_3 $ be a tensor product of three Hilbert spaces
    $ H_1, H_2 $ and $ H_3 $ and let $ \rho_{123} $ be a positive definite operator on $ H_{123}. $ We
    consider the partial traces
    \[
    \rho_2=\tr_2\rho_{123},\quad\rho_{12}=\tr_{12}\rho_{123},\quad\rho_{23}=\tr_{23}\rho_{123}
    \]
    and define
    \[
    k_{123}=k_1\otimes 1_2\otimes 1_3 + 1_1\otimes k_2\otimes 1_3 + 1_1\otimes 1_2\otimes k_3,
    \]
    where $ k_1, k_2 $ and $ k_3 $ are self-adjoint operators on $ H_1, H_2 $ and $ H_3 $ respectively.
    Strong subadditivity of the Wigner-Yanase-Dyson entropy is the condition
    \begin{equation}
    S_p(\rho_{123}, k_{123})+S_p(\rho_2, k_2)\le S_p(\rho_{12}, k_{12})+S_p(\rho_{23}, k_{23}),
    \end{equation}
    where $ k_{12}=k_1\otimes 1_2+1_1\otimes k_2 $ and $ k_{23}=k_2\otimes 1_3+1_2\otimes k_3. $

    Strong subadditivity (SSA) is for the Wigner-Yanase-Dyson entropy a stronger condition than just
    subadditivity (SA). This may be inferred\footnote{The author is indebted to
    E. Lieb and R. Seiringer for providing this argument.} in the following way. We set $ H_2=\mathbf C $ and let
    \[
    \rho_{13}=\sum_i x_i\otimes z_i,\qquad x_i\in B(H_1),\, z_i\in B(H_3)
    \]
    be a positive semi-definite operator on the tensor product $ H_1\otimes H_3 $ and set
    \[
    \rho_{123}=\sum_i x_i\otimes 1_2\otimes z_i.
    \]
    We calculate the partial traces
    \[
    \tr_{13}\rho_{123}=\rho_{13},\quad\tr_{12}\rho_{123}=\tr_1\rho_{123},\quad
    \tr_{23}\rho_{123}=\tr_3\rho_{123}
    \]
    and by setting $ k_2=1_2, $ we obtain $ S_p(\rho_{123},k_{123})=S_p(\rho_{13}, k_{13}) $ and
    \[
    S_p(\rho_2, k_2)=0,\quad S_p(\rho_{12}, k_{12})=S_p(\rho_1,k_1),\quad S_p(\rho_{23}, k_{23})=S_p(\rho_3,k_3).
    \]
    Strong subadditivity thus entails
    \[
    S_p(\rho_{13}, k_{13})\le S_p(\rho_1, k_1)+S_p(\rho_3, k_3),
    \]
    which is subadditivity.

    \section{The WY-entropy is not subadditive}

    We shall demonstrate that the Wigner-Yanase entropy is not subadditive
    (and therefore not strongly subadditive either).

    Consider the tensor product $ H_{12}=H_1\otimes H_2 $ of two Hilbert spaces, each of dimension 2,
    with fixed orthonormal bases $ (e_1,e_2) $ in $ H_1 $ and $ (f_1, f_2) $ in  $ H_2. $ 
    We introduce the positive definite operator $ \rho_{12} $ on $ H_{12} $ with matrix representation
    \[
     \rho_{12} = \begin{pmatrix}
                      7 & 5 & 5 & 6\\
                      5 & 6 & 2 & 5\\
                      5 & 2 & 6 & 5\\
                      6 & 5 & 5 & 7
                      \end{pmatrix}
     \]
     and eigenvalues $ \frac{1}{2}(21+5\sqrt{17}), 4, 1, \frac{1}{2}(21-5\sqrt{17}). $ The 
     coordinates of for example $ \rho_{12}(e_2\otimes f_1) $ are then given by column no. 3 in the matrix representing 
     $ \rho_{12}. $ The positive definite square root $ \rho_{12}^{1/2} $ is represented by
     \[
      \rho_{12}^{1/2} = \begin{pmatrix}
            2 & 1 & 1 & 1\\
            1 & 2 & 0 & 1\\
            1 & 0 & 2 & 1\\
            1 & 1 & 1 & 2
            \end{pmatrix}
      \]
      and have eigenvalues $ \frac{1}{2}(5+\sqrt{17}), 2, 1, \frac{1}{2}(5-\sqrt{17}). $ 
      
      We also consider the operators on $ H_1 $ and $ H_2 $ given by
     
      \[
      k_1=\begin{pmatrix}
                 10 & 1\\
                 1  & 1
                 \end{pmatrix}\quad\text{and}\quad
                 k_2=\begin{pmatrix}
                 1 & 1\\
                 1 & 10
                 \end{pmatrix},
      \]
      and calculate
      \[
      k_1 \otimes 1_2=\begin{pmatrix}
          10 & 0  & 1 & 0\\
          0  & 10 & 0 & 1\\
          1  & 0  & 1 & 0\\
          0  & 1  & 0 & 1
          \end{pmatrix},\qquad
      1_1 \otimes k_2=\begin{pmatrix}
          1 & 1  & 0 & 0\\
          1 & 10 & 0 & 0\\
          0 & 0  & 1 & 1\\
          0 & 0  & 1 & 10
          \end{pmatrix}
      \]
      and      
      \[
      k_{12}=k_1\otimes 1_2 + 1_1\otimes k_2=
      \begin{pmatrix}
                     11 & 1  & 1 & 0\\
                     1  & 20 & 0 & 1\\
                     1  & 0  & 2 & 1\\
                     0  & 1  & 1 & 11
                     \end{pmatrix}.
      \]
      The commutator
      \[
      [\rho_{12}^{1/2}, k_{12}]=\begin{pmatrix}
      0   & 10 & -8 & 0\\
      -10 & 0  & 0  & -10\\
      8   & 0  & 0  & 8\\
      0   & 10 & -8 & 0
      \end{pmatrix}
      \]
      has a square
      \[
      [\rho_{12}^{1/2}, k_{12}]^2=\begin{pmatrix}
      -164 & 0    & 0    & -164\\
      0    & -200 & 160  & 0\\
      0    & 160  & -128 & 0\\
      -164 & 0    & 0    & -164
      \end{pmatrix}
      \]
      and therefore
      \[
      S(\rho_{12}, k_{12})=\frac{1}{2}\tr [\rho_{12}^{1/2}, k_{12}]^2=-328.
      \]
      We calculate the partial traces
      \[
      \rho_1=\rho_2=\begin{pmatrix}
                    13 & 10\\
                    10 & 13
                    \end{pmatrix}
      \]
      with eigenvalues $ (3,23) $ and positive definite square roots
      \[
      \rho_1^{1/2}=\rho_2^{1/2}=\begin{pmatrix}
                                 \frac{\sqrt{3}+\sqrt{23}}{2} & \frac{-\sqrt{3}+\sqrt{23}}{2}\\[0.5ex]
                                 \frac{-\sqrt{3}+\sqrt{23}}{2} & \frac{\sqrt{3}+\sqrt{23}}{2}
                                \end{pmatrix}.
      \]
      The commutators
      \[
      [\rho_1^{1/2}, k_1]=-[\rho_2^{1/2}, k_2]=
      \begin{pmatrix}
      0                                & -\frac{9}{2}(-\sqrt{3}+\sqrt{23})\\[0.5ex]
      \frac{9}{2}(-\sqrt{3}+\sqrt{23}) & 0
      \end{pmatrix}
      \]
      have squares
      \[
      [\rho_1^{1/2}, k_1]^2=[\rho_2^{1/2}, k_2]^2=
      \begin{pmatrix}
            -\frac{81}{4}(-\sqrt{3}+\sqrt{23})^2 & 0\\[0.5ex]
            0                                    & -\frac{81}{4}(-\sqrt{3}+\sqrt{23})^2
            \end{pmatrix}
      \]
      and therefore
      \[
      S(\rho_1,k_1)=S(\rho_2,k_2)=\frac{1}{2}\tr [\rho_1^{1/2}, k_1]^2=-\frac{81}{4}(-\sqrt{3}+\sqrt{23})^2.
      \]
      In conclusion we obtain
      \[
      \begin{array}{rl}
      S(\rho_1, k_1)+S(\rho_2, k_2)-S(\rho_{12}, k_{12})&\displaystyle=-\frac{81}{2}(-\sqrt{3}+\sqrt{23})^2+328\\[2ex]
      &=-725+81\sqrt{69}\\[2ex]
      &\approx-52.1635,
      \end{array}
      \]
      which contradicts the conjecture of subadditivity of the Wigner-Yanase entropy. We report
      without giving a proof that also the Wigner-Yanase-Dyson entropies, for 0<p<1, fail to be
      subadditive.
      
         \section{More general entropies}
         
         The failure of the WYD-entropies $ S_p(\rho, k) $ to be subadditive is not related to the 
         term $ \rho^p $ as one might expect.          
         In a recent paper \cite{kn:hansen:2006:4} we introduced, for each regular monotone metric with
         Morozova-Chentsov function $ c, $ a so called metric adjusted
         skew information $ I^c(\rho, k) $ which is a generalization of the Wigner-Yanase-Dyson skew information.
         The extreme points in the convex set of Morozova-Chentsov functions are given by the functions
   \begin{equation}
   c_\lambda(x,y)=\frac{1+\lambda}{2}\left(\frac{1}{x+\lambda y}+\frac{1}{\lambda x+y}\right)\qquad
   x,y>0,
   \end{equation}
   where $ \lambda\in [0,1]. $ We introduce for each $ \lambda\in(0,1] $ the $ \lambda $-entropy
   \begin{equation}
    E_\lambda(\rho,k)=-I^{c_\lambda}(\rho,k)=-\tr\rho k^2+\tr k\, f_\lambda(L_\rho,R_\rho)k
     \end{equation}
   specified by the function 
   \[
   f_\lambda(x,y)=xy\cdot c_\lambda(x,y)\qquad x,y>0. 
   \]   
   Each $ \lambda $-entropy is a concave function in $ \rho, $
   and it is additive with respect to the aggregation of isolated subsystems \cite{kn:hansen:2006:4}.
   The $ \lambda $-entropies are vanishing for commuting operators $ \rho $ and $ k $ and have therefore no classical
   counterparts. Furthermore, the Wigner-Yanase-Dyson entropy has the following representation
    \begin{equation}\label{integral expression of WYD}
    S_p(\rho, k)=\frac{p(1-p)}{2}\int_0^1 E_\lambda(\rho, k)\frac{(1+\lambda)^2}{\lambda}\, d\mu_p(\lambda)\qquad 0<p<1,
    \end{equation}
    where
    \[
    d\mu_p(\lambda)=\frac{2\sin p\pi}{\pi p(1-p)}\cdot\frac{\lambda^p+\lambda^{1-p}}{(1+\lambda)^3}\,d\lambda
    \]
    is a probability measure on $ [0,1] $ such that $ \int_0^1 \lambda^{-1}d\mu(\lambda)<\infty. $
    
    Although the $ \lambda $-entropies are much simpler than the WYD-entropies, we realize from
    equation (\ref{integral expression of WYD}) that at least one of them must fail to be subadditive,
    and this applies then to all of them because of their similarities.

{\footnotesize


      \vfill

      \noindent Frank Hansen: Department of Economics, University
       of Copenhagen, Studiestraede 6, DK-1455 Copenhagen K, Denmark.}

      \end{document}